\documentclass[twocolumn,aps,pre,showpacs]{revtex4}
\usepackage[latin1]{inputenc}
\usepackage{graphicx}
\usepackage{latexsym}
\usepackage{amssymb}
\usepackage{amsmath}
\begin{document}
\def\d{{\rm d}}
\def\ex{{\rm e}}
\def\v{{\rm v}}
\def\y{{\bf y}}
\def\A{{\bf A}}
\def\O{{\bf O}}
\def\W{{\bf W}}
\def\k{{\rm k}}
\def\r{{\bf r}}
\def\p{{\bf p}}
\def\E{{\bf E}}
\def\bOmega{{\boldsymbol{\Omega}}}
\def\bsigma{{\boldsymbol{\sigma}}}
\def\smalk{{\scriptscriptstyle{\rm k}}}
\def\beq{\begin{equation}}
\def\eeq{\end{equation}}
\font\brm=cmr10 at 24truept
\font\bfm=cmbx10 at 15truept
\title{Concentration fluctuations of large Stokes number 
\\
particles in a one-dimensional random velocity field}
\author{Piero Olla$^1$ and M. Raffaella Vuolo$^2$}
\affiliation{
$^1$ISAC-CNR, and INFN Sez. Cagliari, I--09042 Monserrato, Italy.
\\
$^2$Dipartimento di Fisica and INFN, Universit\'a di Cagliari, 
I--09042 Monserrato, Italy.
}

\date{\today}

\begin{abstract}
We analyze the behavior of an ensemble of inertial particles in a one-dimensional
smooth Gaussian velocity field, in the limit of large inertia, but considering
a finite correlation time for the random field. 
The amplitude of the concentration
fluctuations is characterized by slow decay at large inertia
and a much larger correlation length than that of
the random field. 
The fluctuation structure in velocity space is
very different from predictions from short-time
correlated random velocity fields, with only few particle pairs crossing
at sufficiently small relative velocity to produce correlations.
Concentration fluctuations are
associated with depletion of the relative velocity variance of 
colliding particles.
\end{abstract}

 \maketitle

\section{INTRODUCTION}
Starting from the work of Deutsch \cite{deutsch85}, it has been known for some time now, 
that inertial particles in a random velocity field undergo clustering phenomena.
These behaviors are observed in numerical simulation of turbulence as well
(see e.g \cite{cencini06} and references therein) and are thought to give an important 
contribution to coalescence phenomena, e.g. in the process of rain formation 
\cite{shaw03,falkovich02}. Both in the case of random fields and of real 
turbulence, spatial inhomogeneity of the statistics,
though contributing to particle segregation \cite{brooke94}, 
does not appear to be an essential factor. 

Random velocity fields with various statistical properties,
have been used to model inertial particles in turbulent flows, both in 
the presence of gravity \cite{maxey87,fung03} and in its absence
\cite{elperin96,sigurgeirsson02}. 
Simplified models disregarding in part or in total the spatial 
structure of the velocity field have been introduced as well 
(see e.g. \cite{bec07a,falkovich07}), to cope with the difficulty 
of the analytical treatment of the multi-particle statistics.

In the simplest instance, an 
inertial particle in a turbulent flow is characterized by the relaxation time of
its velocity relative to the fluid: the Stokes time $\tau_S=2/9\ a^2\lambda/\nu_0$, 
where $a$ is the particle radius, $\lambda$ is the ratio of the particle to fluid
density (assumed large) and $\nu_0$ is the kinematic viscosity of the fluid 
\cite{maxey83}. Experimental data 
\cite{fessler94} and numerical simulations \cite{wang93} both indicate that 
clustering is stronger for particles with $\tau_S$ of the order of 
the Kolmogorov time scale. This observation, and the fact
that Stokes times for most atmospheric aerosols of interest lie in this range \cite{shaw03}
have motivated substantial analytical effort in the small Stokes time regime, 
with smooth velocity fields mimicking turbulence at the Kolmogorov scale
\cite{balkovsky01,falkovich02}. 

Some analysis has been carried on also for Stokes times in the turbulent 
inertial range \cite{elperin96,olla02,bec06}, but numerical simulations \cite{bec06a}
indicate that concentration fluctuations do not follow simple scaling rules.

Due to the relative ease of analytical treatment, another regime which 
attracted attention is that of inertial particles in a smooth
random velocity field, with correlation time $\tau_E$ much shorter than the 
geometric scale $r_\v/\sigma_u$ obtained from the correlation length
$r_\v$ and the amplitude $\sigma_u$ 
of the velocity field fluctuations \cite{mehlig04,duncan05}.
However, this is a short-time correlated regime that is very different from 
the situation in realistic turbulent flows: $\tau_E\sigma_u/r_\v=O(1)$.

The analysis of random velocity field models has allowed to identify at least
two clustering mechanisms, expected to be present also
in real turbulence. The first one, originally proposed in 
\cite{maxey87}, is preferential concentration of heavy (light) 
particles in the strain (vortical) regions of the flow. 
This effect has been recently observed also in the case of 
particles with very small inertia \cite{bec06b,bec06c}.
The second mechanism, already present in one dimension (1D)
\cite{wilkinson03,derevyanko06}, is that of the particles catching 
one another in their motion, as they slip with respect to the fluid.
In the case of smooth velocity fields, this leads
to the formation of caustics in the instantaneous concentration field, 
which, in turn, act as back-bones for the clustering process \cite{wilkinson05}.
The clustering itself is a long-time process, which can be described in terms
of the Lyapunov exponents of the particle pair dynamics \cite{wilkinson05,bec07}.

Focusing on the second mechanism, it appears that caustics formation tends to
become maximum for $\tau_S/\tau_E,\tau_E\sigma_u/r_\v\to 1$ \cite{wilkinson06},
and that, at the same time, clustering becomes weaker. For larger $\tau_S$, one expects
that the particles be scattered by the velocity
fluctuations they cross in their motion \cite{abrahamson75}
as if undergoing Brownian diffusion, resulting
in vanishing particle correlations and in clustering destruction.
It is to be stressed that the interest of this limit is by no means academic,
as particles in turbulent flows for which $\tau_S$ lies in the
inertial range (or above), will see smaller vortices precisely in this way.

\vskip 5pt

Purpose of this paper is to understand in detail how and under which conditions
the uncorrelated limit described in \cite{abrahamson75} is achieved.
It will appear that clustering destruction occurs
in a very non trivial way, requiring consideration, among the other 
things, of how particle correlations decay at scales comparable with 
or above $r_\v$. We are going to show that, provided we are away 
from the short correlation time regime $\tau_E\ll\sigma_u/r_\v$, 
clustering at $\tau_S\gg\tau_E$ will not be dominated by the 
small-separation particle pair dynamics (small with respect to $r_\v$)
described in \cite{wilkinson03}.  
Particle pairs remaining close long enough for a Lyapunov exponent
approach to be appropriate, are still present, but 
their contribution to concentration fluctuations is negligible.

The key mechanism for the destruction of concentration fluctuations
will appear to be that, at large $\tau_S/\tau_E$ and finite $\tau_E\sigma_u/r_\v$, 
only particles with 
increasingly small relative velocities (but not small enough for
a local theory on the lines of \cite{wilkinson03,bec07} to be valid) 
have a chance to be correlated.
This is to be contrasted with the picture in \cite{wilkinson06}, of 
correlation fluctuations disappearing as caustics occupy in the above limit
larger and larger portions of space, and with 
the one in \cite{bec07} of saturation to the space dimension of the particle
distribution correlation dimension. Neither in \cite{bec07} nor in 
\cite{wilkinson06}, however, was a
quantitative prediction on clustering decrease provided.

The present analysis will also allow to show that, 
for large $\tau_S/\tau_E$ and rather generic 
(not too large) values of $\sigma_u\tau_E/r_\v$, clustering is associated 
with smaller typical relative inter-particle velocities, compared
to what would be observed in the absence of correlations. 
Thus, there are circumstances under which, clustering may
hinder rather than enhance coalescence phenomena.

This paper is organized as follows. The main definitions and approximations
will be presented in Section II, following by and large the notation of
\cite{wilkinson03,mehlig04,duncan05}. In Section III, the pair particle dynamics
will be analyzed in the large $\tau_S$ limit, identifying the relevant time and 
velocity scales for clustering.
In Section IV, the relation between particle-particle velocity depletion 
and clustering will be established, and heuristic estimates for
the concentration correlation will be obtained, starting from the Fokker-Planck 
equation for the distribution of the relative particle velocity and coordinate.
In Section V, a proof for the dominance of slowly approaching particle pairs
in cluster generation will be provided.
Section VI contains the conclusion.

\section{MODEL EQUATIONS}
Consider an ensemble of inertial particles transported by a 1D
zero mean, Gaussian random velocity field $u(x,t)$ with correlation
\beq
\langle u(x,t)u(0,0)\rangle=\sigma_u^2F(t/\tau_E)g(x/r_\v),
\label{eq1}
\eeq
where $g(0)=F(0)=1$, the function $g$ is assumed to be smooth and to have 
all necessary moments, and $\int_0^\infty F(t)\d t=1$.
The particles are immaterial, so that they can cross without
interaction, and their velocity $v$ is taken to obey the Stokes equation:
\beq
\dot v=\tau_S^{-1}[-v(t)+u(x(t),t)].
\label{eq2}
\eeq
Introduce the
Stokes and Kubo numbers:
\beq
S=\tau_S/\tau_E,
\qquad
K=\sigma_u\tau_E/r_\v
\label{eq3}
\eeq
and choose units so that $\sigma_u=\tau_S=1$; therefore: 
\beq 
\tau_E=S^{-1},
\qquad
r_\v=(KS)^{-1}.
\label{eq4}
\eeq
In the regime $S\gg 1$, the particle velocity will change little on the 
lifetime of a fluid fluctuation and Eq. (\ref{eq2}) could be approximated
by a Langevin equation. This, independently of $K$, i.e. of the random
field being short-time correlated or not.
It is possible to substitute into Eq. (\ref{eq2}):
\beq
u(x(t),t)\to (2\tau_p(v))^{1/2}\xi(t),
\label{eq5}
\eeq
with $\xi(t)$ white noise: $\langle\xi(t)\xi(0)\rangle=\delta(t)$ and 
\beq
\tau_p(v)=
\int_0^\infty\frac{\langle u(x(t),t)u(x(0),0)|v\rangle}
{\langle u^2(x(t),t)\rangle}
\d t
\label{eq5.01}
\eeq
the correlation time of the fluid velocity sampled by a particle moving
at speed $v$. 
In the above formula, the averages are calculated along trajectories,
with $\langle.|v\rangle$ indicating the condition on $v$. The average on the trajectory in 
the denominator, 
due precisely to the presence of concentration fluctuations, does not 
necessarily coincide with the space average 
$\langle u^2\rangle\equiv\sigma_u^2$.

From Eq. (\ref{eq1}), we can put $\tau_p\simeq\tau_E$, provided 
$r_\v/\sigma_v\gg\tau_E$, with $\sigma_v^2$ the particle velocity
variance.  If $\tau_p\simeq\tau_E$, we can estimate from 
Eqs. (\ref{eq2}) and (\ref{eq5}): 
\beq
\sigma^2_v\simeq S^{-1}, 
\label{eq5.1}
\eeq
which is satisfied if $K^2\ll S$. Following the same line
of reasoning (see Appendix), it is possible to show that prescription
problems in the definition of the white noise arise only at $O(K^2/S)$.
Notice that $K^2\ll S$ and $S\gg 1$ imply $\chi=K/S\ll 1$, where $\chi$ is
precisely the force magnitude parameter introduced in \cite{wilkinson03}.
This condition is easily satisfied unless one chooses to work in a
frozen turbulence regime $K\gg 1$. 

Turning to the relative motion of particle pairs, let us introduce
difference variables
$$
\nu=v_2-v_1,
\qquad
r=x_2-x_1,
$$
where 1 and 2 label members of a particle pair, and indicate
$u_{1,2}(t)=u(x_{1,2}(t),t)$.
In the regime $S\gg 1$, starting from Eq. (\ref{eq2}),
we can approximate the equation for the relative motion of particles with the
Langevin equation:
\beq
\dot\nu=-\nu+b(r)\xi,
\qquad
\dot r=\nu.
\label{eq7}
\eeq
For $S\gg K^2$, we have for the noise square amplitude $B(r)=b^2(r)$,
from Eq. (\ref{eq1}):
\begin{eqnarray}
B(r)&=\int_{-\infty}^\infty\langle[u_2(t)-u_1(t)][u_2(0)-u_1(0)]|\nu,r\rangle\d t
\nonumber
\\
\nonumber
\\
&\simeq 4S^{-1}[1-g(r/r_\v)].
\label{eq8}
\end{eqnarray}
The decoupling of the difference variables $(\nu,r)$ from the center of mass 
variables $\frac{1}{2}(x_1+x_2)$ and $\frac{1}{2}(v_1+v_2)$ descends 
basically from the condition $\tau_p(v)\simeq\tau_E$.

\section{TWO-PARTICLE DYNAMICS}
\begin{figure}
\begin{center}
\includegraphics[draft=false,width=7.5cm]{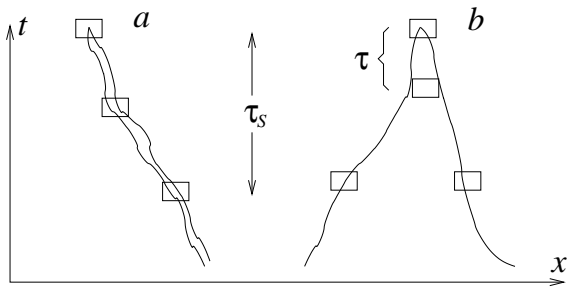}
\caption{
Typical particle pair trajectories for $\epsilon\ll 1$ $(a)$ and
$\epsilon\gg 1$ $(b)$. The exit time to the right is 
$\tau\sim\epsilon^{-1/2}\tau_S$. 
The small rectangles are space-time correlated regions for 
the fluid, of size $r_\v$ and duration $\tau_E$. The time
$\tau_S$ quantifies the particle memory. The particle approach 
in $(a)$ is governed by an $O(\epsilon)$ negative Lyapunov exponent
\cite{wilkinson03}.
}
\label{corrfig0}
\end{center}
\end{figure}

In order to determine how concentration fluctuations are generated, we need
to understand the particle pair dynamics at small separations.  
In the large $S$ limit, we expect that the velocities of colliding
particles be uncorrelated, however, this condition turns out  not to be
automatically satisfied. A condition for the particles to be weakly correlated,
despite being close, is that the Stokes time be much longer than the
time spent at $|r|\lesssim r_\v$, so that the velocity of the two particles
will be the result of many contributions by uncorrelated
regions of the fluid (see Fig. \ref{corrfig0}).
Under these conditions, particles will move ballistically
at scale $|r|\lesssim r_\v$, and they will cross at velocity 
$\nu\sim\sigma_v= S^{-1/2}$, after a time
$\tau\sim r_\v/\sigma_v\sim (K^2S)^{-1/2}$. The weak correlation
condition is therefore $r_\v/\sigma_v\ll\tau_S$, i.e.:
\beq
\epsilon=K^2S\gg 1.
\label{eq9}
\eeq
%
%
The parameter $\epsilon$ is the same as in the theory in
\cite{mehlig04,duncan05}, who considered in detail the 
small $\epsilon$ regime. 
For $\epsilon\ll 1$, the time spent by a pair
of particles at $|r|\lesssim r_\v$ before collision, would be 
longer than $\tau_S$, and the particles would have memory only of 
that last portion of their history, when they were at $|r|\ll r_\v$;
therefore, $\nu$ could not be treated, at the time of crossing, as 
a difference of uncorrelated velocities \cite{note0}.

Notice that the conditions $\epsilon\ll 1$ 
and $S\gg 1$ can be realized only if $K\ll S^{-1/2}\ll 1$, 
corresponding to a short-time correlated random velocity
field. In this regime, $\epsilon$ plays the role of an 
effective Stokes number (see in particular \cite{bec07}), 
in which the place of $\tau_E$ is
taken by $r_\v/(\sigma_u^2\tau_E)$, that is the diffusion 
time of a (non-inertial) tracer across a distance $r_\v$.
Notice also that the condition $\tau_S\gg\tau_E,r_\v/\sigma_u$,
i.e. $KS,S\gg 1$, is not sufficient by itself to guarantee weak correlation,
that in fact is not fulfilled in the range
$S^{-1}\ll K\ll S^{-1/2}$, where $\epsilon\ll 1$.

Back to the large $\epsilon$ range, although most particles will be
weakly correlated, those approaching at speed 
$|\nu|\ll\sigma_v$ may stay at $|r|<r_\v$
sufficiently long to end up being strongly correlated.
It is then important to understand whether an effect like
clustering, which depends on pair-correlation, is due to the 
effect of the many weakly correlated particle pairs, or to the one
of the few strongly correlated ones.

We will answer this question in Sec. V. For the moment,
we identify the important time and velocity scales involved in the
dynamics of slowly approaching particle pairs. We rescale variables:
\beq
\hat t=t/t_\v,
\qquad
\hat \nu=\nu/\nu_\v,
\qquad
\hat r=r/r_\v,
\label{eq10}
\eeq
where
\beq
t_\v=(2\epsilon)^{-1/3},
\qquad
\nu_\v=2^{1/3}\epsilon^{-1/6}\sigma_v
\label{eq11}
\eeq
and $r_\v$ is given in Eq. (\ref{eq4}). In the new variables, at short separations, the
noise amplitude (\ref{eq8}) can be Taylor expanded and Eq. (\ref{eq7}) becomes:
\beq
\dot{\hat\nu}=-(2\epsilon)^{-1/3}\hat\nu+\hat r\xi;
\qquad
\dot{\hat r}=\hat\nu.
\label{eq12}
\eeq
Thus, $t_\v$ and $\nu_\v$ are in some sense the time and velocity scales at the correlated
scale $r_\v$. Notice that the condition for Eq. (\ref{eq12}) to be meaningful as a stochastic
differential equation is $t_\v\gg\tau_E$, i.e. again $\chi=K/S\ll 1$.

To understand better the meaning of $t_\v$ and $\nu_\v$, consider the separation of a 
pair of particles such that $\hat r(0)=0$ and $|\hat\nu(0)|\ll 1$. This is equivalent 
to looking at the approach of a particle pair, with conditions imposed on the velocity at 
the moment of crossing, rather than at a previous  initial time.

In the initial phase, the separating particles will move at almost constant
relative velocity, and $\hat r\simeq\hat\nu(0)\hat t$.
The variation of $\hat\nu$ will be
$\Delta\hat\nu\simeq\hat\nu(0)\int_0^{\hat t}\hat t'\xi(\hat t')\d\hat t'$
so that $\langle \Delta\hat\nu^2\rangle\simeq\hat\nu^2(0)\hat t^3/3$. The initial phase will
last until $\Delta\hat\nu\sim\hat\nu(0)$, i.e. until $\hat t\sim 1$, when $\hat r\sim\hat\nu(0)$.
We then see that $\hat\nu\sim 1$, i.e. $\nu=\nu_{\rm v}$
is the minimum relative
velocity to which the approximation of constant relative velocity at the correlated
scale applies, and $t_\v$ is the associated exit time: $|r(t_\v)|\sim r_\v$.

The following phase will be determined by the linear nature of Eq. (\ref{eq12});
for $t<\tau_S$, the drift $-(2\epsilon)^{-1/3}\hat\nu$ can still be disregarded, and
we expect $\hat\nu(\hat t),\hat r(\hat t)\sim\hat\nu(0)\exp(\hat t)$. 
After a time of the order of $-\ln\hat\nu(0)$, the particles will have moved apart by
$\hat r\sim 1$, i.e. they will be moving out of a correlated region for $u$. In this final
part of the separation process, Eq.  (\ref{eq12}) will cease to be valid, with $\hat\nu$ 
obeying instead $\dot{\hat\nu}\simeq\xi$.
For $\hat \tau_S\gg\hat t\gg -\ln\hat\nu(0)$,
$\nu$ will behave like a Brownian motion: $\hat\nu(\hat t)\sim \hat t^{1/2}$,
while $\hat r$ will undergo a Richardson-like diffusion process: 
$\hat r(\hat t)\sim\hat t^{3/2}$.

In conclusion, we have two groups of particle pairs.
\begin{enumerate}
\item The great majority, for which, at the moment of crossing:
$\nu\sim\sigma_v$; their motion in a correlated region can be
approximated as ballistic and their permanence time
is $\tau\sim\epsilon^{-1/2}\tau_S$.
\item A smaller fraction, for which, at the moment of crossing:
$|\nu|\lesssim\nu_{\rm v}$, with $\nu_{\rm v}\sim\epsilon^{-1/6}\sigma_v$; their
dynamics in a correlated region is determined by fluctuations of $u(x,t)$ at scale
$\lesssim r_\v$
and their permanence time grows logarithmically as $\nu\to 0$: $\tau(\nu)\sim\epsilon^{-1/3}
\ln(\nu_{\rm v}/|\nu|)$.
\end{enumerate}
The fact that for $\epsilon\gg 1$, $t_\v\ll\tau_S$, has the consequence that
only particles meeting at exponentially small relative velocities
$$
\nu_\tau\sim \sigma_v\epsilon^{-1/6}\exp(-{\rm const.}\,\epsilon^{1/3}\tau/\tau_S),
$$
will have a chance to stay within a correlated region for a time
$\tau\gtrsim\tau_S$. An exponentially low level of molecular diffusion 
is thus sufficient to destroy these pairs. Even in the absence of
molecular diffusion, 
however, the relative motion of these pairs is characterized in 
the limit $\tau/\tau_S\to\infty$ by a positive Lyapunov
exponent \cite{wilkinson03} and no long-time contributions to 
clustering are present. The remaining particle pairs can be
shown not to produce singular contributions to the two-particle
probability density function (PDF) for $\nu$ 
\cite{olla07}. We can then estimate the
fraction of particle pairs staying close for a time 
$\tau$, i.e. the fraction of particles
crossing at speed $\nu_\tau$ (provided of course $\tau>t_\v$),
as 
$$
\nu_\tau/\sigma_v\sim
\epsilon^{-1/6}\exp(-{\rm const.}\,\epsilon^{1/3}\tau/\tau_S).
$$ 
This, again, is exponentially small for $\tau\gtrsim\tau_S$ and large $\epsilon$, while
the fraction of particle pairs that are merely not ballistic, is the much larger
value $\nu_\v/\sigma_v\sim\epsilon^{-1/6}$.

The situation in the small $\epsilon$ regime is very different, as 
Eq.  (\ref{eq12}) leads to instantaneous equilibration of the particle velocity:
$\langle\nu^2|r\rangle\sim\sigma_v^2 (r/r_v)^2$ and the ballistic 
particle fraction, associated with caustics formation, disappears 
\cite{wilkinson03,wilkinson06,note01}. Hence,
$\sigma_v^{-2}\langle\nu^2|r=0\rangle\to 0$ and the particles behave for
$\epsilon\ll 1$ 
as a mono-disperse phase (although with velocity not locally equal to that of
the fluid, as it happens instead in the small $S$ limit \cite{balkovsky01}).
At $\epsilon\to 0$, all particle pairs are therefore co-moving.

\section{THE PAIR DISTRIBUTION}
We have seen at the end of Section II that in the regime $S\gg 1$, $S\gg K^2$,
the difference variables $\nu,r$ decouple from the center of mass ones.
This means that at equilibrium, the two-particle PDF 
will be in the form
$$
\rho(v_{1,2},x_{1,2})=\Omega^{-1}\rho
\Big(\frac{v_1+v_2}{2}\Big)\rho(\nu,r),
$$
where $\Omega$ is the length of the domain for $x_{\scriptscriptstyle 1,2}$. 
(Unless ambiguous, we do not use subscripts to identify PDF's referring 
to different stochastic variables).
Multiplying by $N^2$, with $N$ the total number of particles in
the domain $\Omega$ and integrating over $\d v_1\d v_2$,
we obtain the expression for the concentration correlation:
\beq
\langle n(r)n(0)\rangle=\bar n^2\Omega\int\d\nu\rho(\nu,r):=\bar n^2[1+f(r)],
\label{eq14}
\eeq
where $\bar n=N/\Omega$ is the mean concentration. The quantity
$f(r)=\Omega\rho(r)-1$ gives the strength of the
concentration fluctuations.

The difficulty of the problem, compared with the $\epsilon\ll 1$ limit, is the finite 
width in $\nu$ of the PDF $\rho(\nu,r)$. In the $\epsilon\ll 1$ limit, 
small $r$ implied small $\nu$, a long permanence at that particular $r$ and the 
possibility of linearizing the Fokker-Planck equation in $r$ along the lines
of \cite{piterbar02,wilkinson03}. In our case this is not possible, and we must
consider the full Fokker-Planck equation, which,
from Eqs. (\ref{eq7}) and (\ref{eq8}), reads, at steady state:
\beq
\nu\partial_r\rho(\nu,r)-
\partial_\nu[\nu\rho(\nu,r)]=\frac{1}{2}B(r)\partial_\nu^2\rho(\nu,r)
\label{eq15}
\eeq
(see \cite{zaichik03} for another example of kinetic treatment of the particle
pair statistics in turbulent flows).

Taking moments of the Fokker-Planck equation (\ref{eq15}) allows to draw important
conclusions on the connection between clustering and decrease of the relative velocity 
variance. The first two moments of Eq. (\ref{eq15}) read:
\beq
\begin{array}{ll}
\partial_r(\langle\nu|r\rangle\rho(r))=0,
\\
\partial_r(\langle\nu^2|r\rangle\rho(r))+\langle\nu|r\rangle\rho(r)=0,
\end{array}
\label{eq16.01}
\eeq
where $\langle\nu^p|r\rangle\rho(r)=\int\d\nu\nu^p\rho(\nu,r)$. 
We can impose boundary conditions to Eq. (\ref{eq16.01}) at $r\to\infty$,
where the solution, corresponding to uncorrelated particles, is known.
The first moment equation tells us that $\langle\nu|r\rangle\rho(r)$
is constant; but, at $|r|\gg r_\v$ correlations are absent; therefore 
$\rho(r\to\infty)=\Omega^{-1}$ and $\langle\nu|\infty\rangle=0$;
thus $\langle\nu|r\rangle\rho(r)=0$ for all $r$. Substituting
into the second one, we get $\partial_r(\langle\nu^2|r\rangle\rho(r))=0$,
and, using again 
$\rho(r\to\infty)=\Omega^{-1}$ and $\langle\nu^2|\infty\rangle:=\sigma_\nu^2
=2S^{-1}$, we get:
\beq
\sigma_\nu^{-2}\langle\nu^2|r\rangle=[1+f(r)]^{-1}.
\label{eq16.02}
\eeq
In other words, clustering in 1D is necessarily associated with velocity variance
decrease. Notice that this result rests only on the applicability of
the Fokker-Planck equation (\ref{eq15}), which requires $S\gg 1,\chi=K/S\ll 1$,
but is otherwise independent of $\epsilon$.

An idea of how $f(r)$ goes to zero as $\epsilon\to\infty$ can be obtained treating
the function $g$ in Eq. (\ref{eq8}) as a perturbation. This is motivated by
the observation (see previous section) that most particles spend very little
time at distances $|r|<r_\v$, and by the assumption (verified in \cite{olla07})
that the remaining particles produce only a correction to $\rho(r)$. 
To lowest order, the particles are therefore independent random walkers with 
velocity correlation time $\tau_S$.
A further simplification is obtained assuming that the scale of variation
of $\rho(\nu,r)$ remains $\sigma_v=S^{-1/2}$, as in the unperturbed case.
We can then write $\rho\simeq \rho_0+\rho_1$, with 
\beq
\rho_0(\nu,r)=\Omega^{-1}\rho_0(\nu)=\frac{S^{1/2}}{2\pi^{1/2}\Omega}\exp(-\frac{S\nu^2}{4})
\label{eq16.1}
\eeq
the solution of Eq. (\ref{eq15}) with $g(r/r_\v)$ set to zero,
and estimate $\nu\partial_r\rho_1\sim S^{-1/2}\partial_r\rho_1$,
$\partial_\nu(\nu\rho_1)-(1/2)B_0\partial_\nu^2\rho_1\sim\Gamma\rho_1$
for some constant $\Gamma$. Substituting into Eq. (\ref{eq15}):
\beq
[S^{-1/2}\partial_r+\Gamma]\rho_1
\sim B_0g(r/r_\v)\rho_0.
\label{eq16.2}
\eeq
A new length $S^{-1/2}$, the distance travelled by a particle in a 
Stokes time, thus enters the problem.
On the scale of $S^{-1/2}$,
the term $g$ in Eq. (\ref{eq16.2}) behaves like a Dirac delta and
we find:
\beq
\rho_1(\nu,r)\sim\Omega^{-1}\epsilon^{-1/2}\exp(-\Gamma S^{1/2}|r|)\rho_0(\nu).
\label{eq17}
\eeq
Substituting into Eq. (\ref{eq14}) would lead to 
concentration fluctuations with $O(\epsilon^{-1/2})$ variance and 
$O(S^{-1/2})$ correlation length.

\section{THE VELOCITY STRUCTURE}
The analysis carried on in the previous section did not answer the 
question of which particle pairs are responsible for the production 
of the concentration fluctuations at large $\epsilon$. We are going to show here that
concentration fluctuations are indeed produced by slowly approching
particle pairs. 

We prove that the contrary is impossible and
assume therefore that the dominant contribution is due to
particles approaching ballistically at scale $r_\v$, which
spend a time $\tau\sim\epsilon^{-1/2}\tau_S$ at that scale.
In order to show that the assumption of dominance of ballistic 
particles cannot be true, it is sufficient to calculate 
$\rho(r)=\int\rho(\nu,r)\d\nu$ as the stationary solution 
of the evolution equation
\beq
\rho(r;0)=\int\d r'\d\nu'\rho(r;0|\nu',r';-T)\rho(\nu',r';-T),
\label{eq17.1}
\eeq
where $\rho(\nu,r;t)$ is a generic 
non-equilibrium PDF
and $\rho(r;0|\nu',r';-T)\equiv\rho(r(0)$=$r|\nu(-T)$=$\nu',r(-T)$=$r')$ is a transition PDF
for the dynamics described by Eqs. (\ref{eq7}-\ref{eq8}).
If we take $\tau\ll T\ll\tau_S$ and $|r|\lesssim r_\v$, for most particles [i.e. for typical
values of $\nu(-T)$], we will have:
\beq
|r(-T)|\simeq |r(0)-\nu(-T)T|\gg r_\v,
\label{eq17.2}
\eeq
which descends from the fact that, for $T\ll\tau_S$ and $\nu\sim\sigma_v$, $\nu$ can 
be approximated as constant. But Eq. (\ref{eq17.2}) tells us that the dominant 
contribution to the integral in Eq. (\ref{eq17.1}) for $\rho(r;0)=\rho(r)$ (the
equilibrium PDF), is from values of $r'$ for which 
$\rho(\nu',r',-T)=\rho(\nu',r')\simeq\rho_0(\nu',r')$,
where $\rho_0$ is the spatially homogeneous PDF of Eq. (\ref{eq16.1}).
For the ballistic particle pairs (which we are assuming to dominate the statistics),
we can calculate the correction to ballistic motion, writing $r(t)=r_0(t)+r_1(t)$, where, 
using Eq. (\ref{eq7}):
\beq
\begin{array}{ll}
r_0(t)&=r(-T)+\nu(-T)(t+T),
\\
r_1(t)&\simeq \int_{-T}^t\tau\d\tau b(r_0(\tau))\xi(\tau).
\end{array}
\label{eq18}
\eeq
Using Eq. (\ref{eq18}), we can express the transition PDF $\rho(r(0)|\nu(-T),r(-T))$
in Eq. (\ref{eq17.1}) in terms of the equivalent PDF 
$\rho(r_1(0)|\nu(-T),r(-T))$ and, using the first of Eq. (\ref{eq18}), in terms of
the conditional PDF $\rho(r_1(0)|\nu(-T),r_0(0))$. Substituting into Eq. (\ref{eq17.1}),
with $\rho(r;0)=\rho(r)$ and $\rho(\nu',r';-T)\simeq\rho_0(\nu',r')
=\Omega^{-1}\rho_0(\nu')$,
we obtain:
\begin{eqnarray}
\rho(r)=
\Omega^{-1}
\int\d r_1(0)\d\nu(-T)
\rho_0(\nu(-T))
\nonumber
\\
\times
\rho(r_1(0)|\nu(-T),r_0(0))
\Big|_{r_0(0)=r-r_1(0)},
\label{eq19}
\end{eqnarray}
where use has been made of the relation
$\rho(r)=\int\d r_0(0)\d r_1(0)\rho(r_0(0),r_1(0))
\delta(r_0(0)+r_1(0)-r)$.
We can Taylor expand Eq. (\ref{eq19}) around $r_0(0)=r$ and we obtain
$$
\rho(r)\simeq\frac{1}{2\Omega}\partial_r^2\langle r_1^2(0)|r_0(0)=r\rangle,
$$
where the average is on $\nu(-T)$ (in terms of  $\rho_0$) and $\xi(\tau)$.
Using Eqs. (\ref{eq18}) and (\ref{eq8}), we can write:
$$
\rho(r)\simeq\frac{1}{2\Omega}\partial_r^2\langle\int_{-T}^0\tau^2\d\tau B(r+\nu\tau)\rangle,
$$
where the remaining average is on $\nu$;
setting $r=0$, we see 
from Eq. (\ref{eq8}) that the time integral is dominated by 
$\min(T,r_\v/|\nu|)$ and we obtain the final result
\beq
\rho(r=0)\sim\Omega^{-1}r_\v^{-2}B(0)\langle (r_\v/|\nu|)^3|\,|\nu|>r_\v/T\rangle,
\label{eq20}
\eeq
where the conditional average to right hand side is peaked at $|\nu|\sim r_\v/T$ and diverges 
for $\epsilon,T/\tau_S\to\infty$.  

Thus, ballistic pairs do not dominate the dynamics.
A calculation taking into account small velocities in consistent way \cite{olla07},
in turn, would lead to convergent expressions for $\rho(\nu,r)$ and
allows to show that the contribution to concentration fluctuation is concentrated at 
$\nu\sim\nu_\v\sim\epsilon^{-1/6}\sigma_v$, as could have been guessed from the 
analysis in Section III.

\section{CONCLUSION}
The large Stokes number limit of inertial particles in a random
velocity field is plagued by several subtleties. As pointed out in
\cite{mehlig04,duncan05}, when this limit is reached, the
particle dynamics becomes dependent on a new parameter
defined in terms of the Stokes and Kubo numbers $S$ and
$K$ [see Eq. (\ref{eq3})], namely: $\epsilon=K^2S$.
This becomes apparent in the case of short-time correlated
velocity fields, such as the Kraichnan model \cite{kraichnan94},
in which $K=0$, and the role of the Stokes number is played indeed by
$\epsilon$ [from Eq. (\ref{eq3}), $S$ is trivially equal to
infinity]. Thus, as pointed out in \cite{bec06}, the large
Stokes number asymptotics is really the large $\epsilon$ one
of Eq. (\ref{eq9}).

The point of this paper is that, for $\epsilon\gg 1$,
the concentration dynamics is not governed by particle pairs 
staying close long enough for a local theory on the lines of
\cite{balkovsky01,piterbar02,wilkinson03} to work.
While, for small $\epsilon$,
the particle phase is locally mono-disperse
in velocity \cite{note2}, 
for $\epsilon>1$, the particle velocity distribution 
has finite width and only very few such 
almost co-moving particle pairs exist (see discussion
at the end of Section III). This property is expected
to hold also in more than 1D.

Small relative velocities still appear to be important:
as suggested in \cite{abrahamson75}, the particles 
behave like a gas in thermodynamic equilibrium, but the correlations 
originate, rather than from random collisions, from particle
encounters at small relative velocities, as made clear by the analysis
leading to Eq. (\ref{eq20}). A qualitative argument in Section III,
confirmed by the analysis in \cite{olla07}, suggests 
that the only particle pairs that 
have a chance to be correlated are those travelling at 
relative velocity $\sim\epsilon^{-1/6}\sigma_v$, that is 
the minimum for ballistic relative motion at scale $r_\v$.
However, these velocities  are still $O(\exp(\epsilon^{1/3}))$ larger than those
producing the long time dynamics needed for clustering in 
the $\epsilon\ll 1$ limit.

The concentration 
fluctuation variance appears to decay rather slowly,
like $\epsilon^{-1/2}$ at large $\epsilon$ [see Eq. (\ref{eq17})].
This decay rate could be interpreted as the product of
the concentration fraction $\sim\epsilon^{-1/6}$ 
of the particles staying close long enough for their 
relative velocity to be modified in some way,
and their permanence time 
$\sim\epsilon^{-1/3}$ 
(in units $\tau_S$)
at separation $|r|<r_\v$
[see Eq. (\ref{eq11})]. 

This mechanism of concentration fluctuation production
leads to an important modification of the relative
velocity distribution. As particle clusters are
the result of particle pairs with small relative
velocities, the total velocity PDF will be
weighed more to small relative velocities. 
This results in a depletion of the relative 
velocity variance, quantified in the exact 
relation (\ref{eq16.02}), that becomes substantial at
$\epsilon\sim 1$. It is to be ascertained whether
this property is preserved in more than 1D.

Summarizing, we have the following situation: extrapolating
the small $\epsilon$ theory of 
\cite{wilkinson03,bec07,wilkinson06} to $\epsilon\gtrsim 1$, 
predicts that clusters become less and less singular,
making eventually a transition from fractal
to space-filling objects. Extrapolating the opposite way
around, the present
large $\epsilon$ theory predicts
''residual'' (not singular) concentration 
fluctuations, becoming of the order
of the mean concentration as $\epsilon\to 1$, 
due to more and more particle pairs 
contributing to the fluctuations.
It is reasonable to expect that the situation at 
$\epsilon\sim 1$ will be a combination of the two pictures.

In \cite{falkovich02,wilkinson06}, it was suggested
that clustering is not central in determining
collision rates of relevance e.g. in  rain formation,
as instead are caustics formation and a finite width 
of the velocity distribution. If the present picture
continues to hold in more than 1D, it would result in 
an even stronger statement, that is, at large $\epsilon$, clustering
and modifications to the velocity distribution 
work against one another. This is in some way analogous 
to the situation at $\epsilon\to 0$,
of clustering dominance accompanied by
caustics disappearance and vanishing of collision rates
\cite{wilkinson03} (see also \cite{zhou98}).

Another feature of concentration fluctuations at large $\epsilon$,
is the strong scale separation between their decay length 
and the correlation length of the fluid: the first
is by a factor $\sim\epsilon^{1/2}$ larger than 
the second [see Eq. (\ref{eq17})]. The concentration fluctuation
decay length is basically the distance travelled
by a particle in a Stokes time and is independent
of the fluid correlation structure. Taking 
into account also the $\epsilon^{-1/2}$ variance decay,
smaller vortices may thus contribute
in a non negligible way
to the concentration fluctuations of particles with 
Stokes time in the turbulent inertial range.

\appendix
\section{FINITE STOKES NUMBER CORRECTIONS}
The Langevin approximation to Eq. (\ref{eq2}) provided by (\ref{eq5}) is the lowest order
in an expansion in inverse powers of the Stokes number $S$. Let us calculate the next order.
Given initial conditions at $t=0$, Eq. (\ref{eq2}) can be integrated to give:
$$
v(t)=v(0)\ex^{-t}+\int_0^t\d t'\ex^{t'-t}
u(x(t'),t'),
$$
and we can write $x(t)=\bar x(t)+\tilde x(t)$ with
$$
\bar x(t)=x(0)+v(0)(1-\ex^{-t})
$$
and
$$
\tilde x(t)=\int_0^t\d t'(1-\ex^{t'-t})u(x(t'),t').
$$
Following an approach similar to \cite{maxey87}, 
the perturbation expansion is carried on with respect to $\tilde x$ and the first correction
to $v(t)$ is given by, for $t\ll\tau_S=1$:
$$
\begin{array}{ll}
v(t)=v(0)(1-t)+\int_0^t\d t'u(\bar x(t'),t')
\\
+\int_0^t\d t'\int_0^{t'}\d t''(t'-t'')u'(\bar x(t'),t')u(\bar x(t''),t'')
\end{array}
\eqno(A1)
$$
where $u'(x,t)=\partial_xu(x,t)$ and $\bar x(t)\simeq x(0)+v(0)t$. Taking
$t\gg\tau_E$, we find that the second piece to right hand side of Eq. (A1) 
behaves like a Wiener increment; using Eq. (\ref{eq5.01}):
$$
\Big\langle\Big[\int_0^t\d t'u(\bar x(t'),t')\Big]^2\Big\rangle=2\tau_p(v(0))t
$$
and we recover Eq. (\ref{eq5}). Notice that the substitution $x(t')\to\bar x(t')$ 
avoids a definition of the noise in implicit form and $\xi(t)$ in Eq. (\ref{eq5}) 
is automatically defined in the It\^o prescription \cite{gardiner}.

The last term in Eq. (A1) contains a drift correction in the form, from Eq. (\ref{eq1}):
$$
\begin{array}{ll}
\langle u'(\bar x(t'),t')u(\bar x(t''),t'')\rangle
=-(KS)^2v(0)(t'-t'')
\\
\times
\langle u(v(0)(t'-t''),t'-t'')u(0,0)\rangle. 
\end{array}
$$
Equation (A1) takes then the final form, in the limit $t\to\d t$:
$$
\d v=-[1+\gamma(v)]v\d t+(2\tau_p(v))^{1/2}\d w
\eqno(A2)
$$
where 
$$
\gamma(v)=(KS)^2\int_0^\infty\d t\ t^2\langle u(vt,t)u(0,0)\rangle.
\eqno(A3)
$$
and $w(t)=\int_0^t\xi(\tau)\d\tau$ is the Wiener increment. The noise amplitude 
is defined up to $O(\d t^{1/2})$ terms associated with the fluctuating part of 
the second line of Eq. (A1).

Thus, the first order correction to the Langevin equation for $v(t)$ is 
an $O(K^2/S)$ renormalization of the Stokes time, whose sign depends on the
profile of the correlation function for $u(x(t),t)$: a strictly positive 
correlation corresponds to a decrease of the Stokes time, while oscillations 
in the correlation may lead to an increase. This would lead to a decrease
(an increase) of the drift induced on the particle by an external force.

Notice that the drift correction $\gamma$ provided in Eq. (A3) does not necessarily
coincide with what would be expected from a change from Stratonovich to It\^o 
prescription \cite{gardiner}. For instance, adopting for the random field correlation 
the explicit expression $g(x)=\exp(-x^2/2)$, would lead to twice what would be
obtained 
interpreting the substitution $x(t')\to \bar x(t')$ in 
$\int_0^t\d t'u(\bar x(t'),t')$ as a Stratonovich to It\^o prescription change.
Indicating
by $\d_S$ the Stratonovich differential and writing $b(v)=(2\tau_p(v))^{1/2}$:
$$
b(v)\d_S w=\frac{1}{2}b(v)b'(v)\d t+b(v)\d w,
$$
we find in fact, from Eq. (\ref{eq5.01}): $\frac{1}{2}b(v)b'(v)=-\frac{1}{2}\gamma v$.



\begin{thebibliography}{99}

\bibitem{deutsch85} J.M. Deutsch ''Aggregation-disorder transition induced
by fluctuating random forces'', J. Phys. A: Math. Gen. {\bf 18}, 1449 (1985)

\bibitem{cencini06} M. Cencini, J. Bec, L. Biferale, G. Boffetta, A. Celani,
A.S. Lanotte, S. Musacchio and F. Toschi ''Dynamics and statistics of heavy
particles in turbulent flows'', JoT {\bf 7}, 1 (2006)

\bibitem{shaw03} R.A. Shaw ''Particle-turbulence interactions in atmospheric
clouds'', Annu. Rev. Fluid Mech. {\bf 35}, 183 (2003)

\bibitem{falkovich02} G. Falkovich, A. Fouxon and M.G. Stepanov, ''Acceleration 
of rain initiation by cloud turbulence'', Nature {\bf 419}, 151 (2002)

\bibitem{brooke94} J.W. Brooke, T.J. Hanratty, J.B. Mc Laughlin ''Free-flight
mixing and deposition of aerosols'', Phys. Fluids {\bf 6}, 3404 (1994)

\bibitem{maxey87} M.R. Maxey ''The gravitational settling of aerosol particles
in homogeneous turbulence and random flow fields'', 
J. Fluid Mech. {\bf 17}, 174 (1987)

\bibitem{fung03} J.C.H. Fung and J.C. Vassilicos ''Inertial particle segregation by 
turbulence'', Phys. Rev. E {\bf 68}, 046309 (2003)

\bibitem{elperin96} T. Elperin, N. Kleeorin and I. Rogachevskii ''Self-excitation of
inertial particle concentration in turbulent fluid flows'', Phys. Rev. Lett.
{\bf 77}, 5373 (1996)

\bibitem{sigurgeirsson02} H. Sigurgeirsson and A.M. Stuart ''A model for preferential
concentration'', Phys. Fluids {\bf 14}, 4352 (2002)

\bibitem{bec07a} J. Bec and R. Ch\'etrite ''Toward a phenomenological approach to the
clustering of heavy particles in turbulent flows'', e-print nlin/0701033

\bibitem{falkovich07} G. Falkovich, S. Musacchio, L. Piterbarg and M. Vucelja
''Inertial particles driven by a telegraph noise'', e-print nlin/0703055

\bibitem{maxey83} M.R. Maxey and J.J. Riley ''Equation of motion for a small rigid particle in a
non-uniform flow'', Phys. Fluids {\bf 26}, 883 (1983)

\bibitem{fessler94} J.R. Fessler, J.D. Kulick and J.K. Eaton ''Preferential 
concentration of heavy particles in a turbulent channel flow'', Phys. Fluids
{\bf 5}, 3742 (1994)

\bibitem{wang93} L.-P. Wang and M.R. Maxey ''Settling velocity and concentration 
distribution of heavy particles in homogeneous isotropic turbulence'', J. Fluid 
Mech. {\bf 256}, 27 (1993) 

\bibitem{balkovsky01} E. Balkovsky, G. Falkovich and A. Fouxon ''Intermittent 
distribution of inertial particles in turbulent flows'', Phys. Rev. Lett. {\bf 86},
2790 (2001)

\bibitem{olla02} P. Olla ''Particle transport in a random velocity field 
with Lagrangian statistics'' Phys. Rev. E {\bf 66}, 056304 (2002)

\bibitem{bec06} J. Bec, M. Cencini and R. Hillerbrand ''Clustering of heavy particles
in the inertial range of turbulence'', 
e-print nlin/0606038

\bibitem{bec06a} J. Bec, L. Biferale, M. Cencini, A. Lanotte, S. Musacchio and F. Toschi
''Heavy particle concentration at dissipative and inertial scales'', 
e-print nlin/0608045

\bibitem{mehlig04} B. Mehlig and M. Wilkinson ''Coagulation by random velocity fields as
a Kramers problem'', Phys. Rev. Lett. {\bf 92}, 250602 (2004)

\bibitem{duncan05} K. Duncan, B. Mehlig, S. \"Ostlund and M. Wilkinson ''Clustering by
mixing flows'', Phys. Rev. Lett. {\bf 95}, 240602 (2005)

\bibitem{wilkinson03} M. Wilkinson and B. Mehlig ''Path coalescence transition and
its applications'', Phys. Rev. E {\bf 68}, 040101(R) (2003)

\bibitem{bec06b} J. Bec, L. Biferale, M. Cencini, A.S. Lanotte and F. Toschi
''On the effects of vortex trapping on the velocity statistics of tracers and
heavy particles in turbulent flows'', e-print nlin/0604059

\bibitem{bec06c} J. Bec, L. Biferale, G. Boffetta, A. Celani, M. Cencini, A. Lanotte, 
S. Musacchio and F. Toschi ''Acceleration statistics of heavy particles in turbulence'',
e-print nlin/0508012

\bibitem{derevyanko06} S.A. Derevyanko, G. Falkovich, K. Turitsyn and S. Turitsyn
''Explosive growth of inhomogeneities in the distribution of droplets in a 
turbulent air'' JoT (in press) e-print nlin/0602006

\bibitem{wilkinson05} M. Wilkinson and B. Mehlig ''Caustics in turbulent aerosols'',
Europhys. Lett. {\bf 71}, 186 (2005)

\bibitem{bec07}  J. Bec, M. Cencini and R. Hillerbrand, ''Heavy particles in 
incompressible flows: The large Stokes number asymptotics'', Physica D {\bf 226}, 11
(2007)

\bibitem{wilkinson06} M. Wilkinson, B. Mehlig and V. Bezugly ''Caustic activation
of rain showers'', e-print cond-mat/0604166

\bibitem{abrahamson75} J. Abrahamson ''Collision rates of small particles in a vigorously
turbulent fluid'', Chem. Eng. Sci. {\bf 30}, 1371 (1975)

\bibitem{note0} This would be true also if the particles approached with speed $\sigma_v$, like
in the $\epsilon\gg 1$ case; it is then necessarily true also in the present case, where
$|\nu|\ll\sigma_v$ (see discussion at the end of the Section).

\bibitem{note01} A simple alternative way to see how caustics disappear 
for $\epsilon\to 0$: 
from Eq. (\ref{eq12}),
$\langle\hat\nu^2|\hat r\rangle\sim \hat r^2\epsilon^{1/3}$, while the velocity required
for a ballistic flight to $\hat r=0$ is
$\hat\nu\sim \hat r/\hat\tau_S\sim \hat r\epsilon^{-1/3}$. The probability of such a
flight would be: 
$\sim\exp[-\hat\nu^2/(2\langle\hat\nu^2|\hat r\rangle)]
\sim \exp(-c/\epsilon)$, with $c$ a constant.


\bibitem{piterbar02} L.I. Piterbar ''The top Lyapunov exponent for stochastic flow
modelling the upper ocean turbulence'', SIAM J. Appl. Math. {\bf 62}, 777 (2002)

\bibitem{zaichik03} L.I. Zaichik and V.M. Alipchenkov ''Pair dispersion and preferential
concentration of particles in isotropic turbulence'', Phys. Fluids {\bf 15}, 1776 (2003)
 
\bibitem{zinn} J. Zinn-Justin, {\it Quantum Field Theory and Critical Phenomena}, 4th edition
(Clarendon Press, Oxford 2002)

\bibitem{olla07} P. Olla and M.R. Vuolo, 
'' Perturbation theory for large Stokes number particles in random velocity fields''
e-print arXiv:0801.3204


\bibitem{kraichnan94} R.H. Kraichnan, ''Anomalous scaling of a randomly 
advected passive scalar,'' Phys. Rev. Lett. {\bf 72}, 1016 (1994)

\bibitem{note2} Apart near caustics, where the distribution is a superposition of mono-disperse
jets.


\bibitem{zhou98} Y. Zhou,, A.S. Wexler and L.-P. Wang ''On the collision rate of small
particles in isotropic turbulence. II. Finite inertia case'', 
Phys. Fluids {\bf 10}, 1206 (1998)

\bibitem{gardiner} C.W. Gardiner, {\it Handbook for stochastic methods}, third edition 
(Springer NY, 2004)




\end{thebibliography}
\end{document}